# Top Down Approach: SIMULINK Mixed Hardware / Software Design

Youssef atat[1,3] and Mostafa Rizk[1,2]

[1] Computer Science Department, Lebanese University,
Beirut, Lebanon
[2] Electronics Department, Telecom Bretagne,
Technople Brest Iroise 29238 Brest France
[3] Institut des Sciences Appliquées et Economiques (ISAE)
Beirut, Lebanon

**Abstract**
System-level design methodologies have been introduced as a solution to handle the design complexity of mixed Hardware / Software systems. In this paper we describe a system-level design flow starting from Simulink specification, focusing on concurrent hardware and software design and verification at four different abstraction levels: System Simulink model, Transaction Simulink model, Macro architecture, and micro architecture. We used the MP3 CodeC application, to validate our approach and methodology.

## 1. Introduction

Technical and no technical constraints are taken in consideration when designing an embedded system. The complexity of the application, the difficulties of testing and validation of the whole hybrid system and the limits of methodologies present the technical constraints. The time to market factor presents the no technical constraint. So, in order to deal with these complex embedded systems and to meet the more severe time-to-market constraints, we need new design methods. Researchers are classified into two categories: microelectronic researchers which try to integrate more and more transistors on the same chip, and design flow researchers which try to develop new methodologies of embedded system design or new tools that facilitate the process of conception.

These methodologies are based on high level design and have emerged as promising approach to cope with this challenge [6], [7]. Indeed, another problem appears in the increasing of the abstraction level, the problem known as application/architecture adequacy.

The designers have proposed many solutions to solve this problem, many frameworks are developed like transactional environments between application development and architecture synthesis [1], [2], or many design tools are developed in order to improve embedded system performances [3], [4], [5].

Consequently, the current challenge is the easy progress of the functional modeling to implementation model, under strong constraints of quality and time design.

In this paper we address the system level design issues for MPSoC design by providing a new approach to bridge the gap between the algorithm design and architecture design.

We make use of a Simulink-SystemC MPSoC design flow that focuses on mixed hardware and software refinement from a high-level Simulink model [14].

The remainder of this paper is organized as follows. In Section 2, we present the proposed system-level design methodology. Four levels of abstraction will be defined, allowing progressive and systematic design and refinements of MPSoC starting from Simulink. We present, also in section 3, the assistance design tools used to reduce the gaps between the various abstraction levels in an effective way. The experimental results for multimedia application will be illustrated through all stages of the design flow to prove the availability and the efficiency of the proposed approach, in Section 4 and draw some conclusions in Section 5.

## 2. Methodology and Design Flow

Our integrated MPSoC design environment shown in Figure 1 consists of algorithm design flow, Architecture design flow, hardware synthesis flow, software design flow, and verification flow for the entire MPSoC system and hardware modules. The main feature of our configuration is that the tool flow is based on: Simulink/Matlab environment [8] for algorithm design and exploration, IP block-based design system [9][10][11], called "Macrocell builder", for hardware synthesis, HW/SW interface synthesis tool called "Roses" [12]. These tools and others tools used in our design flow will be explained later. In our flow, the designer describes the Matlab description not only as a

"specification or modeling language," but also as a final "implementation language" (i.e., bit true coding in Matlab or C language using Simulink S-functions [8]) since we want to avoid double coding for simulation and synthesis. Initially, a designer verifies a pure algorithm for MPSoC in Matlab language. Then, he divides it into software (in S-functions-based Simulink blocks) and hardware descriptions (in Matlab). In hardware design flow, designers use behavioral IPs, as will be explained later. The Macrocell builder system transforms the Matlab description into Synthesizable VHDL RTL. The equivalence between the Matlab and RTL descriptions are guaranteed by our ADCM-based verifier [10]. The software design flow takes a concurrent process specification as an input. Software synthesis constructs an FSM for each process, and then merged them into a sequential software program with the scheduler (written with C code). FSM construction is performed by a dynamic loop scheduling (DLS) and direct state insertion into the original code. The entire system containing hardware and software parts is simulated in our system level cosimulation technique. It can simulate the behavior and cycle-level model. With this cosimulation technique, hardware and software can be designed simultaneously. In our Simulink-based design environment, all simulation models for hardware and software are described in SystemC language. Cycle-level hardware simulation models are generated by Macrocell builder system.

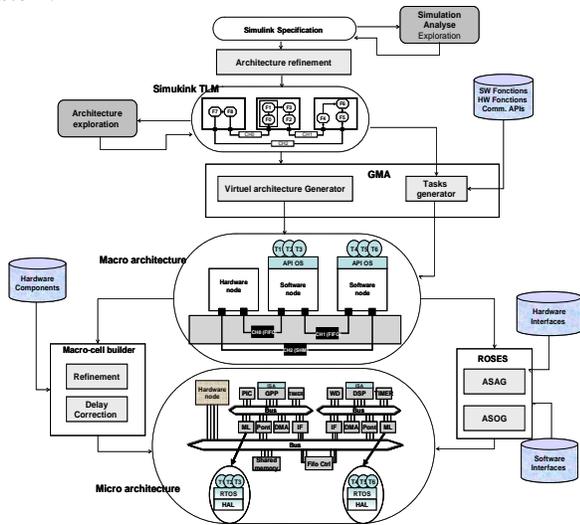

Figure 1. Simulink-based MPSoC Design

## 2.1 Automatic Code Generation for MPSoC Platform through Several Abstraction levels

The proposed system level methodology progresses systematically through successive refinements according to several stages and levels. The goal is to design HW/SW MPSoC architecture and mapping the application algorithms on MPSoC platform. Thus to achieve this goal, it is necessary to reduce the large existing gap between the algorithm design and the architecture description. In this design flow, functional description is a specification which aims initially the signal processing applications. The Simulink environment performs functional modeling and exploration of application algorithms through iterative simulations. The Roses tool [12] permits the HW/SW interfaces design of MPSoC architecture starting from abstract representation netlist of architecture described in architecture description language called COLIF [13]. Regarding this great existing difference between the functional level and the architecture level, it is relevant to define a new transactional level model (TLM) in Simulink as an intermediate level in order to reduce this gap. This transactional model allows to combine algorithm and architecture. In theory, the TLM makes possible the interaction between the components of calculation and communication described at different abstraction levels by hiding the calculations, communications, and their interfaces details or by clarifying them according to the abstraction level decided for each object. To obtain these goals, abstraction rules are defined to model TLM architecture in Simulink.

The MPSoC is formed through a complex heterogeneous component assembly. In order to simplify the design process, we need to employ a certain number of abstraction models or intermediate descriptions. Thus we can define through our design flow four different abstraction levels Abstract specification model for functional algorithm specification in Simulink, TLM in Simulink for combining algorithm and architecture description, the macro-architecture model for HW/SW interfaces design, and the micro-architecture model (implementation level, i.e., synthesizable RTL for hardware parts, and target C code for software parts).

These intermediate models cut out the whole design in several small design stages. Each stage has a specific design objective. As these models can be simulated and estimated, the result of each one of these design stages can be validated independently. Modeling the system at different description levels is necessary, for three reasons: 1) to provide a suitable design level to validate some implementation details for a model representing a system or a part of a system. 2) to provide a sufficient speed of simulation to validate and correct the model. 3) to facilitate and automate the implementation and the transposition of an application on various platforms. The purpose of the prototyping process worked out by this methodology is the automation of the implementation of signal processing applications on a heterogeneous MPSoC platform. Consequently several tools are used to establish this automation. The validation of the system is done through progressive multilevel simulations by specific co-simulation tools for each environment.

## 2.2 Abstract functional Simulink model

The abstract functional model is created through assembly of the pre-modeled parameterized blocks written in high-level language (i.e. Matlab, C/C++). The choice of block parameter values is made by the system designer in order to satisfy a trade-off between signal quality and implementation constraints. The abstract functional model describes the applications with a coarse granularity (Figure 1). With this order of magnitude, each block forming part of the application can be validated independently of the system and can be re-used. This order of magnitude must allow the explicit representation of parallelism and pipeline. We will consider only discrete model with a global clock. The functional model in this case is sensitive to an abstract clock which synchronizes the operation of the system. In order to precisely represent global data and control dependencies, we have selected a subset of Simulink for modeling high throughput multimedia applications (i.e., predefined blocks, delay blocks, User-defined-blocks, if-else structure blocks, for-loop structure blocks, and data/control links blocks). The delay-block is used if an algebraic loop is present in the system. It represents the algorithmic delay that is intrinsic to the algorithm of a block and is independent of CPU speed. It is expressed in terms of the number of samples by which a block's output lags behind the corresponding input. The delay Z-k implies that the output is delayed from the input by k samples. The predefined blocks (Addition AND, Demux ...) are executed in zero algorithmic delay. The user-defined blocks called "S-functions" are used in our abstract functional model to design algorithms in C code and/or in Matlab. The s-function blocs written in Matlab will be used by the hardware design tool Macrocell builder to produce the hardware dedicated portion of the system. The S-functions written in C, are for software synthesis tool. The most adequate configuration model is defined to facilitate functional modeling and multi-threading code generation.

## 2.4 TLM architecture in Simulink

It is an intermediate transaction level in Simulink used to facilitate the procedure of SoC design and implementation and to fill the existing difference between the functional model (neither hardware, nor software) and the virtual architecture model. At this level, the homogeneous functional units will be assigned with software tasks or with hardware IPs. The partitioning choices are taken on this level: Several inter-connected functional units are amalgamated to form a software task or hardware IP. Several functional units and tasks are gathered in a subsystem to form a software node. Several functional units and hardware IPs are gathered in a subsystem to form a hardware node (Figure 1). About the communication, the inter-connected lines at the functional level are simple. These lines are modified in a subsystem to define several types of communication topologies between the hardware\software nodes. This topology can be point-to-point, multipoint, network communication. Moreover, these inter-connected blocks can be implemented on this abstraction level in the Simulink environment to validate several types and communication protocols. The Simulink simulator ensures the mechanism of the communication and the scheduling which remains always implicit. The importance of this stage is to allow the architecture exploration in the Simulink environment.

## 2.3 Macro architecture model

This abstraction level describes the architecture of the system by completely excluding all the details related to realization. The system is carried out in the form of a virtual architecture formed by virtual components. A virtual component can implement software tasks or hardware functions but without any precise characteristics on the type of the component or its internal structure. The virtual component can be carried out like a hardwired architecture, or programmable processor, or a DSP, by carrying out a software function. The description of the system on the virtual architecture level is a whole of such virtual components working jointly and being communicated via explicit and abstract channels. The channels of communication use transactional primitives defined by standard TLM, to represent only the transfer or data synchronization process between the virtual components without any information on the communications protocol implementation. These primitives (API) added to the tasks; correspond to a vision who abstracts the OS layer of the software application (Figure 1). The communications mechanism is done through a data request or a data block achieved in only one transaction. Time is indicated like "time spent" rather than an event by clock. The communication on this abstraction level is no temporal "untimed", whereas the part of calculation is roughly precise in the time "approximately timed" by estimating the execution time on the specific components. The model of the system on this level is however exploitable by using tools and methods for the performance analysis.

## 2.5 Micro architecture model

The systems described on this level have information on the communication protocols implementation and the hardware and software components used in subsystems of the application. On the micro-architecture level (Figure 1) the communication is precisely on the level cycles, while the hardware components remain functional and approximate compared to the execution time. The transactional virtual channels of the preceding level are replaced by transfer channels. A transfer channel is precisely on the level cycles with signals represented by a variable instantiation. The data is transferred according to the precise order of a protocol from bus. The abstraction comes primarily from the bonds which permit to describe a whole of physical signals by only one logical channel. The primitives used are usually Write (addr,dated,ctrl) and Read(addr,dated,ctrl). Addressing is explicit in order to designate the concerned corresponding element which is in fact a memory or a register. The software part on this level is represented by the code implementation, the operating system and HAL layer (Hardware Abstraction Layer). Through the design, the software can thus be simulated by a processor simulator. All the software on this level is compiled and assembled with the instruction set of the target processor. The validation of the system model on this level is carried out by the hardware and the software cosimulation. The simulation model of the system on this level is composed:

— Software subsystems: represented by instruction set simulators of the processor and program in machine language containing the software application.

— Hardware subsystems: represented by behavioral components using of the functional models of bus, and a cosimulation bus. A cosimulation bus is used because several simulators are used to validate the system. The cosimulation bus is a simulators adapter which allows the communication between the various simulators.

## 3. Tools used to generate an MPSoC platform

The design tools used in the proposed design flow are articulated around an intermediate model, COLIF architecture description language [13]. These tools are detailed in the following sections.

3.1 Macroarchitecture generator

A discontinuity and a large gap exist between the functional level and the architectural level while designing an MPSoC. In our design flow, this appears by the hole, present between the model specified in Simulink and the model at the macro architecture level described in COLIF. The generator of the macro architecture permits to automate the passage from the Simulink model to the virtual architecture model in COLIF. This reduces the large hole and establishes continuity between the two models, while accelerating the procedure of the multiprocessor system on chip design. The input of the tool is the Simulink TLM architecture, the output is a macroarchitecture model (behavioral in SystemC and architecture netlist in COLIF).

The macroarchitecture generator is composed of two parts: the architecture generator, and the SystemC behavior generator.

Colif architecture generator consists of parsing the Simulink input into a database tree which stores all required information about the system. It generates from the tree a Colif description keeping the hierarchy. It creates for each element in the transaction Simulink model a correspondent element in the virtual architecture model. It also generates a set of template parameters files according to this architecture. These parameters are specified by the designer to be imported after this and attached to the Colif description. These parameters represent the semantic of the virtual architecture. Figure illustrates the generation of the three basic elements (the module, the port and the Net) of the COLIF virtual architecture starting from the similar elements which define TLM Simulink architecture.

SystemC behavior generator: The tasks at the transactional Simulink level are included in a software node represented by a subsystem having the 'SW_' prefix in its name. These tasks are modeled in Simulink in several ways. It can be formed by an amalgamation of several blocks in a subsystem having the name preceded by the prefix 'TASK_' or they are formed by individual blocks. The latter, in their turn can be a predefined library blocks or S-functions modeled in C language. The tool generates the behavior of the Task by different way depending from the TLM Simulink model: (1) Direct Transformation of Simulink S-functions in SystemC (2) Creation of SystemC from a predefined block in the Simulink library (3) Merge several Simulink blocks SystemC task. Figure 3 illustrates the amalgamation of several blocks in transactional Simulink description to produce a task in SystemC. The library functions of the "F0 (),F1()" have the same names of "F0(), F1()" Simulink blocks. The generation of APIs is done by identifying the protocol type in each module port of the COLIF virtual architecture.

3.2 Automatic generation of HW/SW interfaces

ROSES, is made of several tools, it allows the automatic generation of the HW/SW interfaces of heterogeneous MPSoC [12]. The HW/SW interface (Figure 4) is a communication adapter between the tasks carried out by the processor and the hardware resources. More precisely the software interface permits the code implementation to reach the hardware resources of the processor and the hardware interface permits the processor to reach the communication network. The hardware interfaces taken into account by ROSES concern the processors interfaces, the global memory, and the hardware dedicated components. Roses starts from the macroarchitecture level (i.e., Colif architecture netlist and SystemC behavior description) and generates HW/SW glue (hardware glue in RTL-VHDL and software glue in C and assemblers languages) allowing the HW/SW components assembly of MPSoC system (including cores, hardwired IPs, memories, buses, application software codes, operating system, drivers, …).

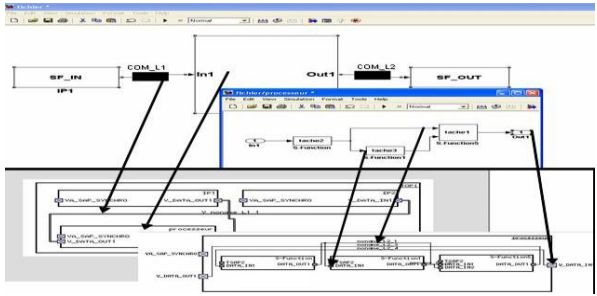

Figure 2. Basic elements generation from Simulink TLM

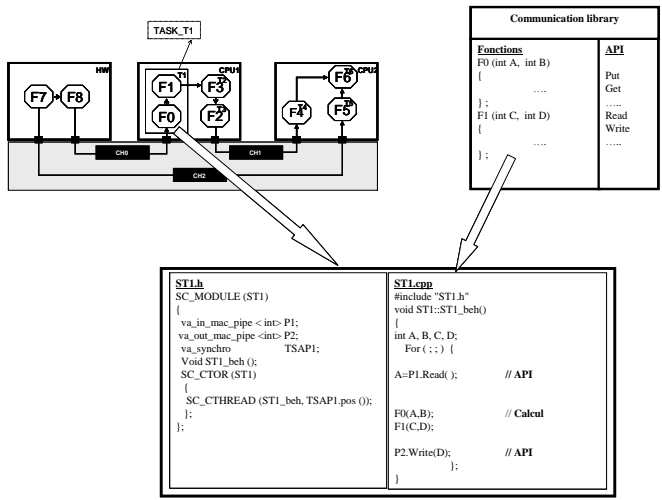

Figure 3. SystemC generation from a set of Simulink blocks

### 3.3 Macro cell builder

The Macro cell builder tool [9], allows the refinement of the functional models, to generate a hardware dedicated portion of MPSoC in synthesizable HDL. This refinement is carried out by correcting the delays due to the behavior difference between functional model and RTL model. These two models are designed starting from two libraries of basics predefined and parameterized blocks (Functional and RTL libraries). The implemented methodology tool is described in Figure below. After the generation of virtual architecture by GMA tool. The Macro cell builder intervenes to treat the Macro hardware blocks (hardware nodes) which are formed by several functional IPs. The refinement process preserves same architecture and replaces each functional IP by the corresponding RTL IP.

Consequently, to generate RTL hardware architecture, the values of the IPs parameters are extracted from the functional model and are used to define the RTL IPs, described in a synthesizable hardware language. Then the delay is corrected by the insertion of registers or by the insertion of a finite state machine FSM. The RTL generated model is adapted for the logical synthesis.

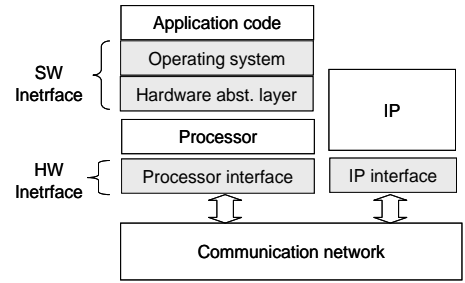

Figure 4. Hardware\Software interfaces generation

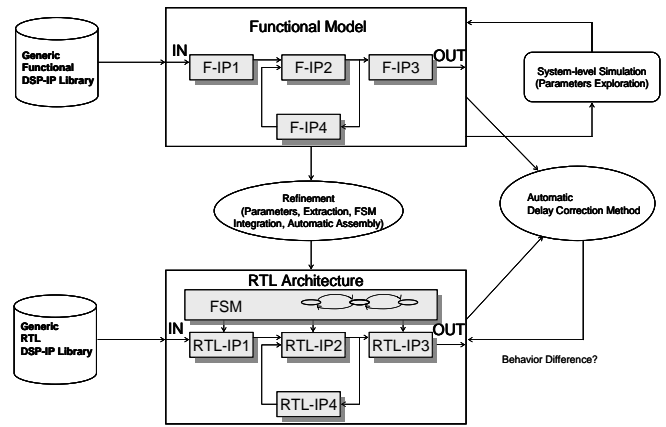

Figure 5. Macrocell builder HW dedicated portion

## 4. MP3 CODEC APPLICATION

We propose to implement the MP3 decoder with two functional IPs. We encapsulate the blocks (tasks) in a subsystem of Simulink, except for the first read unit of MP3 file and the last block of writing in the audio file. The system, in partitioning stage under Simulink is well represented in Figure 6.

The stage which comes now is the passage to the macro architecture in Colif, by using the GMA mapping tool described previously. The structure of MP3 application into Colif is represented in Figure 7. The generation of the hardware interfaces is made starting from a library of components and adapters. Consequently, the internal architecture, the module adapters, and the channels adapters of the processor are present. Afterwards, the software interfaces are generated. The size of the generated OS for this application which is formed of 16 tasks is 25KB. Table 1 summarizes the simulation and co-simulation time at the different abstraction level through the proposed design flow. The Simulink description including algorithm exploration and a system level simulation of MP3 decoder lasts 3 times less than the macro architecture specification and validation. The difference is due to the simulink environment; facilities of debug and the abstraction of the communication via the macro architecture level description. Considering the simulation speed at 4 abstraction levels, the simulation time of the simulink model is 50 time faster than macro architecture.

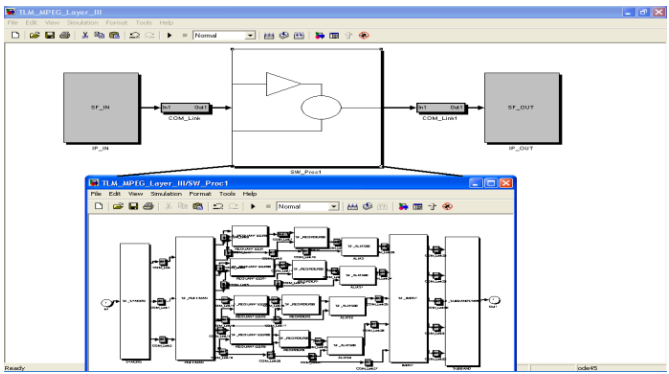

Figure 6. MP3 partitioning application at TLM level

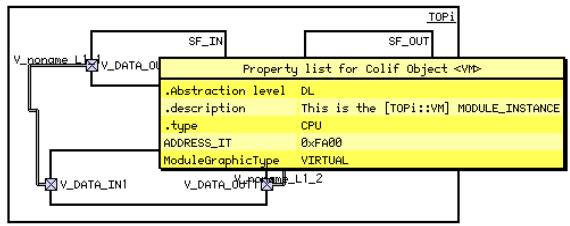

Figure 7. Colif Macroarchitecture of MP3 decoder

The difference is mainly due to the fact that the description at macro-architecture level is more complex and the communication is detailed. The validation with Native OS simulation using UNIX host is 100 times faster than the simulation at ISA level (instruction set architecture). This difference demonstrates clearly the benefits of the validation phase at the different levels. In each level a part of the SoC is validated progressively. At the simulink level, the application algorithm is validated. At the macro architecture level the communication and the interaction between tasks are validated. At the OS level the native simulation validate the application written on the top of an OS. And finally the whole architecture is validated with a co-simulation at cycle accurate with an ISS.

## 5. Conclusion and perspective

In this paper, we have presented a methodology and framework for system design, validation and fast prototyping of multimedia application for multiprocessor System-On-Chip. The present approach takes into consideration system-level-specification, multi-level validation, algorithm exploration, refinement and prototyping generation. Our design flow combines different languages and tools, such Matlab\Simulink, Colif, Interfaces generators of Roses, to reach RTL level. We focused on the tool for the macro architecture generation. The developed tool fills the gap between Simulink ––like Simulation and validation environment of the applications ––and the architectural representation of the applications. We have also presented the design and validation of the MPEG layer III decoder using our design flow. We target an architecture formed by one processor and 2 functional IPs. The future work aims to apply the same methodology to design a more complex architecture that contains several processors. The successful design and result of this case study demonstrated the effectiveness of our approach.

Table 1. Simulation time required to decode 4s of MP3 sound

| MP3 Codec Models | Simulation time |
|---|---|
| TLM Simulink model | 5s |
| Macro-architecture model | 56s |
| Micro-architecture model (Native OS Simulation-based ) | 35mn |
| Micro-architecture model (RTL and Instruction Set Architecture) | 55h |

**Youssef Atat** received the engineer degree in computer communication engineering from the Lebanese University, Beirut Lebanon in 2002, and the MS degree in electronics from the Institut National des Sciences Appliqués (INSA), Rennes, France in 2003, and PHD Degree in microelectronic from the "Institut National Polytechnique de Grenoble" (INPG), Grenoble, France, in 2007. Since 2008 he is an associate professor at the Lebanese University and on the head of regional center of the "Institut des Sciences Appliquées et économies" (ISAE-Lebanon) attached to the "Conservatoire National des arts et des métiers" (CNAM-Paris). His current research interests include system-on-chip design methodology, especially, design and exploration of application-specific multiprocessor architectures, performance estimation, and on-chip communication architectures.

**Mostafa Rizk** received the two Masters degrees in Biomedical and electronics from the Lebanese University Beirut, Lebanon in 2009 and 2010 respectively. He is with the electronics department of the telecom Brest, France. His current research interests are design of application-specific multiprocessor architectures, Non instruction Set Computer, performance, estimation.